Viewpoint

# Examining the Implementation of Digital Health to Strengthen the COVID-19 Pandemic Response and Recovery and Scale up Equitable Vaccine Access in African Countries

Olufunto A Olusanya[1*], MPH, MD, PhD; Brianna White[1*], BSc, MPH; Chad A Melton[2], MSc; Arash Shaban-Nejad[1], MPH, PhD

[1]Department of Pediatrics, The University of Tennessee Health Science Center, Memphis, TN, United States
[2]Bredesen Center for Interdisciplinary Research and Graduate Education, The University of Tennessee, Knoxville, TN, United States
[*]these authors contributed equally

**Corresponding Author:**
Olufunto A Olusanya, MPH, MD, PhD
Department of Pediatrics
The University of Tennessee Health Science Center
50 N Dunlap
Memphis, TN, 38103
United States
Phone: 1 9792043683
Email: oolusan1@uthsc.edu

## Abstract

The COVID-19 pandemic has profoundly impacted the world, having taken the lives of over 6 million individuals. Accordingly, this pandemic has caused a shift in conversations surrounding the burden of diseases worldwide, welcoming insights from multidisciplinary fields including digital health and artificial intelligence. Africa faces a heavy disease burden that exacerbates the current COVID-19 pandemic and limits the scope of public health preparedness, response, containment, and case management. Herein, we examined the potential impact of transformative digital health technologies in mitigating the global health crisis with reference to African countries. Furthermore, we proposed recommendations for scaling up digital health technologies and artificial intelligence–based platforms to tackle the transmission of the SARS-CoV-2 and enable equitable vaccine access. Challenges related to the pandemic are numerous. Rapid response and management strategies—that is, contract tracing, case surveillance, diagnostic testing intensity, and most recently vaccine distribution mapping—can overwhelm the health care delivery system that is fragile. Although challenges are vast, digital health technologies can play an essential role in achieving sustainable resilient recovery and building back better. It is plausible that African nations are better equipped to rapidly identify, diagnose, and manage infected individuals for COVID-19, other diseases, future outbreaks, and pandemics.

*(JMIR Form Res 2022;6(5):e34363)* doi: 10.2196/34363

## Introduction

The COVID-19 pandemic is a novel coronavirus caused by a pathogen known as SARS-CoV-2 [1]. The virus, which is rapidly transmitted via respiratory droplets, was first identified in late 2019 as a potential public health threat in Wuhan, China. Since its identification, the SARS-CoV-2 has had immense adverse impacts across the globe, taking the lives of more than 6 million individuals [1]. The disease often causes mild to moderate respiratory illness; however, older people and those with comorbidities such as hypertension, diabetes, obesity, asthma, or HIV are more likely to develop severe complications as a result of the infection [1]. As the disease rapidly progressed, the World Health Organization (WHO) declared a global health emergency in January 2020, leading to public health preventive and restrictive measures such as lockdowns, stay-at-home orders, quarantine, shelter-in-place orders, curfews, mask mandates, hand hygiene practices, and social distancing, which were swiftly implemented to facilitate emergency response and case containment [1].

After the WHO's global health emergency declaration, Africa saw its first wave of COVID-19 infections, with the earliest cases reported in Egypt in early February 2020. A few weeks later, subsequent reports of infection were made in Algeria, Cameroon, Morocco, and Nigeria [2,3]. Although the prevalence of Africa's first wave seemed paradoxical given the comparatively lower-than-expected reported case numbers, the second wave of infection was more severe [2,4]. By December 31, 2020, only 36 countries had implemented at least 5 of the public health preventive measures; that is, international travel restrictions; workplace, school, and university closures; stay-at-home requirements, etc [2]. While the number of cases fell as the region began to emerge from its second wave, officials warned that the resurgence of the SARS-CoV-2 was inevitable [5]. The rapid spread of the Delta (B.1.617.2) and Omicron (B.1.1.529) mutants have largely driven the most recent surge in cases. As of March 27, 2022, COVID-19 had resulted in an overall estimated prevalence of 11.32 million cases and 250,948 deaths in Africa [6]. At the time, these accounted for 2.4% of cases and 4.1% of deaths recorded globally. Moreover, over 250,000 COVID-19 confirmed cases have been recorded in each of these 10 out of 55 African countries, respectively [6]. There is scientific consensus that COVID-19 vaccines' acceptance and uptake are effective in limiting the transmission of SARS-CoV-2 as well as reducing disease severity, hospitalizations, and deaths [7]. Considering that some higher-income countries have begun to offer booster shots, it is concerning that only 15.3% of the population in African countries has been fully vaccinated [7].

Without rapid tracing and containment of new cases, Africa could see a prodigious increase in morbidity and mortality. The reemergence and resurgence of the SARS-CoV-2 continues to have enormous negative impacts on psychosocial, economic, and health care systems. This warrants an immediate call to action for innovative, state-of-the-art digital health interventions that expedite a resilient recovery and mitigate this and future pandemic crises in African countries. Accordingly, our current investigation examined the potential impact of transformative digital health in alleviating the spread of SARS-CoV-2. In addition, we highlighted digital health technologies adopted worldwide to tackle SARS-CoV-2 transmission and proposed recommendations for scaling up digital health and artificial intelligence while also facilitating equitable vaccine access within the African continent.

## Application of Digital Health and Artificial Intelligence in Preparedness and Rapid Responses

The all-encompassing framework of digital health incorporate components such as mobile health (mHealth) and medical mobile apps, health information technology, smart devices, wearable sensors, wireless medical devices, personalized medicine, telemedicine, and telehealth, which have revolutionized health care systems [8,9]. Moreover, digital health incorporates the application of artificial intelligence and machine learning to enable gathering, management, integration, mining, and interpretation of enormous heterogeneous data or information that includes biomarkers (e.g., disease prevalence and physical activity) and socio-markers (e.g., zip code–level or neighborhood characteristics, education, income level, housing quality, and nearest health facility) [10].

Consequently, we propose that the availability and access to real-time health data through digital health technologies (DHTs) in Africa could facilitate (1) the application of rapid predictive monitoring systems, geographic information, and dashboards for disease surveillance, data visualization, and health decision-making; for example, vaccine and masks mandates, lockdowns, and quarantine measures; (2) good governance or policy decision-making and an inclusive regulatory or legal framework to deter excess bureaucracy and minimize the depletion of scarce resources; (3) leveraging of mHealth and social media analytics to obtain data that assess public sentiments, opinions, or information gaps and enable the creation and dissemination of tailored messages that address mis- or disinformation, vaccine hesitancy, and vaccine inequity; (4) comprehensive educational or training programs that expand digital literacy, health information–seeking behaviors, and precision health education or promotion. Overall, in this pandemic era, the fast-tracking adoption of digital technologies to collect, share, and analyze socio-behavioral and health data could further transform disease preparedness and response, mitigate the spread of infections, and optimize health care delivery services; for example, vaccine delivery, contact tracing, case containment, and management [11-13].

## Social Determinants of Health Risk Factors and Other Barriers to Rapid Recovery

The re-emergence of mutant variants has heightened awareness for the reassessment and reinforcement of current surveillance methods to evaluate behavioral risk factors and ensure protection against future variants [14]. However, given that Africa's health system is particularly vulnerable, efficient policy rollout and implementation may be quite difficult. Poor access to health care, long hospital wait times, ill-equipped and severe shortage of health care professionals, scarcity of personal protective equipment and medical commodities, socioeconomic devastation, as well as disruptions to HIV, tuberculosis, and vaccination programs exacerbate the current COVID-19 burden [5,15-19]. Moreover, lack of government policy and legal framework, inadequate infrastructure, poor maintenance culture, and costly computing resources negatively impact DHT implementation [19]. Inadequate coordination and knowledge for a large-scale rollout of digital health setups, low-level utilization of electronic health records (EHRs), restricted access to reporting systems, and miscalculation of the disease burden owing to limited testing capacity also limit widespread preventive efforts; for example, vaccinations and contact-tracing [20,21]. Besides, some African countries are disproportionately impacted by widespread poverty, low digital literacy levels, internet shortages, food insecurity, climate, and environmental injustices, as well as war, conflicts, and terrorism. Despite the barriers to gathering and retrieving indigenously owned digitally generated data for integration into artificial intelligence systems, the alternate use of externally generated data (from other countries) ultimately creates a predicament [22].

## Digital Health Best Practices That Have Tackled Outbreaks and the Current Pandemic

Digital health technologies have played a central role in the mitigation of the pandemic globally. Even with the myriad of challenges, Africa has seen a number of its own health care advances. In recent years, digital health has come to play a vital role in Africa's public health disease containment efforts. The Africa Centres for Disease Control and Prevention (Africa CDC) and Prevention, a centralized unit for public health surveillance and emergency responses, was created in 2017 [3]. During Nigeria's first COVID-19 surge, social media platforms, text and electronic messaging, telecommunication media, and an artificial intelligence–interactive voice response systems were leveraged by the Nigeria CDC to rapidly and simultaneously disseminate accurate information and debunk myths [23,24]. In Morocco, South Africa, Sierra Leone, and Tunisia, drones were used to enforce compliance with lockdown and social distancing guidelines [25]. Whereas, in Rwanda, the District Health Information Software system presented real-time surveillance data that enabled timely contact-tracing and case management of infected individuals [26].

Similarly, Africa has benefited from a wide range of health emergency responses related to tuberculosis, HIV, and malaria [15]. Ghana, Kenya, and Tanzania adopted an integrated cloud-based mHealth smart reader system to rapidly interpret and transmit diagnostic malaria test results to patients [27]. Other open-access web-based platforms used for comprehensive displays of data have been proposed to condense the plethora of malaria surveillance systems that are in use and scattered across a number of sectors. This integrated and organized approach collects all necessary tools for accurate data retrieval, management, and disease forecasting into one digital space for interoperability [28,29].

Most notably, many African countries have relied on contact-tracing driven by digital health technologies in their fight against Ebola outbreaks since 2014 [23,30]. Consequently, effective contact-tracing strategies originating from networks on Ebola, HIV, Tuberculosis, and Lassa fever have revolutionarily changed how data are collected and transmitted to regional public health centers within Africa. Traditional contact-tracing methods (i.e., paper reports) have been expanded and decentralized to incorporate the use of smartphones, other mobile apps, cell phone tower data, and geospatial mapping [5,21,27] so that real-time data (temperature readings and symptom questionnaire responses) become immediately available as opposed to the lengthy process of collecting paper reports [27,30].

It is widely known that surveillance of population movement and interaction is important in controlling disease transmission, prompting many countries to place a focus on DHTs capable of recording both movement and relevant environmental biomarkers. These location-based biomarkers range from fine particulate matter in the air to descriptive statistics detailing local access to green space and public transportation [31]. Nongenetic biomarkers such as these are reflective of a population's "exposome," a public health concept demonstrating the connection between environmental pressures and overall health status [32]. Over time, external pressures from environmental determinants influence an individual's biological index, making disease development and progression unique [32]. For instance, as mis- or disinformation, politicized debates on vaccine safety, and socio-contextual barriers worsen vaccine hesitancy, public health surveillance methods (e.g., topic modeling, semantic network analysis, and sentiment analysis) can be used to assess and improve the population's "digital exposome" [33,34]. Furthermore, these methods can be used in conjunction with informatics from innovative technologies—for example, wearable sensor devices, smartphone-based sensors, environmental hyperspectral and remote sensing campaigns, and geolocation technologies—to investigate exposome complexities and aid in region- or site-specific, culturally sensitive mitigation plans for African populations.

Moreover, best practices from other countries can also be adapted in Africa within an appropriate sociocultural context. Being the first country affected by the COVID-19 pandemic, China mobilized the use of smartphone apps to keep record of all human movement via public transportation during major outbreaks [13]. This app allowed users to quickly determine exposure during travel and expedite quarantine measures for disease containment [13]. Further, artificial intelligence systems were employed to expedite diagnostic reading time for investigative imaging tools. This resulted in better clinical decision-making and forecast of workforce needs, thus limiting the strain on health care systems [13]. Moreover, Israel used digital vaccine certificates through smartphone apps, termed the "green pass," which allowed admission to certain places; for example, social events for those who are vaccinated [35]. Icelandic scientists developed an app where COVID-19 symptom data (fever and cough) were entered for physicians' monitoring. If symptoms were severe or life-threatening, patients were admitted to a hospital, thus limiting unnecessary hospital visits and health care workers' burnout or exhaustion [36]. In New Zealand, the Ministry of Health developed and distributed the NZ COVID Tracer app, which allowed users to keep digital diaries of their travels and receive alerts following exposed to COVID-19 [37]. In Colombia, the CoronApp was adopted by the National Institute of Health in conjunction with the Columbian Field Epidemiology Training Program (FETP). This digital tool enabled access to FETP teams with data for approximately 5.5 million active users, thus facilitating rapid coordination with local health officials to identify new cases as well as inform preparedness and response decisions [38]. In the United States, 32 states developed mobile apps for contact-tracing using location, Bluetooth, Google or Apple, or DP-3T. Depending on the app's design, capabilities and functionalities varied from a public health official's phone call to travel diaries or notifications on a smartphone.

## Recommendations to Prioritize Scaling up of Transformative Digital Health Technologies in Africa

The application of digital health initiatives has shown considerable success globally. Although digital health technologies have played a large role in facilitating better health outcomes over the past decade, their implementation has become substantially pivotal and timely during the current COVID-19 pandemic. Using existing technologies in combination with those that have demonstrated success in other nations could prove to be advantageous for African countries to manage health crises, facilitate pandemic response or recovery, and promote vaccine uptake. Following the post–COVID-19 era, African countries can be better positioned to overhaul their health care systems to advance the quality of health care, increase cost-effectiveness, bolster health care infrastructure and resources, as well as alleviate the already limited health care workforce. Herein, we examined the potential impact of transformative digital health technologies in alleviating the pandemic's adverse effects on health and health care systems. Furthermore, we proposed recommendations for scaling up digital health technologies and artificial intelligence–based platforms to tackle transmission of the SARS-CoV-2 and enable equitable vaccine access through the following: (1) application of rapid predictive monitoring systems, geographic information, and dashboards for disease surveillance, data visualization, and health decision-making; for example, vaccine and mask mandates, lockdowns, and quarantine measures; (2) good governance or policy decision-making and an inclusive regulatory or legal framework to deter excess bureaucracy and minimize the depletion of scarce resources; (3) leveraging of mHealth and social media analytics to obtain data that assesses public sentiments, opinions, or information gaps and enable the creation and dissemination of tailored messages that address mis- or disinformation, vaccine hesitancy, and vaccine inequity. In addition, the application of DHTs could advance (4) comprehensive educational or training programs that facilitate digital literacy, health information–seeking behavior, and precision health education or promotion.

Despite all the challenges in eHealth investment [39], Africa has taken giant steps forward in using artificial intelligence in precision health [40-42] and precision agriculture [43]. The observation of success stories in using advanced analytics and digital health solutions by researchers and scientists in several African countries, shows a significant hope in term of feasibility of the suggested innovations. Although advancements in African technology and digital health solutions are sure to bolster disease response, investments in policy, and data governance are arguably most essential in moving the needle forward. Web-based open-source, modular digital health platforms with a user-friendly graphical interface and disease surveillance tools that provide accurate, timely depictions of disease case totals, geographic hotspots, vaccine distribution mapping, and supply chain management are crucial to intercepting the spread of SARS-CoV-2 [15,21]. Real-time systematic collection and analysis of data could be useful in guiding governmental support of public health responses, such as vaccine delivery and education, tailored to areas or regions that are more susceptible and with high infectivity rates. Moreover, the development of disease projection and prediction models using machine learning techniques could inform general evidence-informed policy. These integrated disease surveillance tools should be leveraged to not only fight against COVID-19 but also influence future health policy, decision-making, and program implementation.

Notably, a major challenge in some African countries is the lack of adequate infrastructure for internet connectivity, power or electricity supply, and EHR management, which are needed for any real-time application for disease surveillance and monitoring. While countries such as Kenya, Libya, and Nigeria have considerably good internet coverage (approximately 80% on average), others including Madagascar, South Sudan, and Western Sahara have minimal coverage (less than 10%) [44]. African nations with high internet access are suitably positioned to leverage applications for contact-tracing, disease surveillance, data visualization, and vaccine distribution [13,45]. For harder-to-reach or rural areas (without internet access), satellite internet devices and offline digital health strategies can be adopted to collate, integrate, and analyze population data. It is imperative for countries with densely populated city centers and limited internet connectivity (e.g., Egypt and South Africa) to improve and stabilize their capacity for implementation. A positive future outlook is the internet coverage that has rapidly increased by approximately 12,000% in 2020-2021) within Africa [44,45].

Given the complexity of system implementation, it is also essential that barriers such as weak infrastructural investments, lack of funding, absence of regulatory or legal framework, low cost-effectiveness, and poor maintenance and governance are addressed [46,47]. To tackle cyber threats, data insecurity, legal and ethical issues (e.g., end user's consent and privacy, and data ownership), and other unforeseeable related events, countries' regulatory and legal frameworks should be reassessed, reinforced, as well as socioculturally and contextually responsive. Moreover, these legal processes ought to protect individuals' legitimate rights without jeopardizing the implementation of innovative technologies. Additionally, the use of noncomplex, interoperable, integrated, and synchronized methodologies to implement digital health systems could deter excess bureaucracy and duplication of efforts as well as minimize the depletion of scarce resources. While Africa's mobile phone usage falls below that of the global average, its increasing prevalence is adequate to harness and ramp up digital health initiatives [44]. Driven by population migration to cities and urban renewal, it is expected that cell phone use will increase by approximately 1 billion users, including 50% who will have mobile internet access by 2025 [44,48]. This growing use of smartphones and wearable devices or sensors should be leveraged to collate and manage real-time heterogeneous data. Moreover, there is potential for large-scale performance and connectivity emanating from multilateral collaborations, cross-border mHealth programs, and private-sector partnerships to ensure cost-effectiveness and sustainability [49].

In addition to concerns surrounding structural connectivity and mobile phone usage, mistrust in public or government institutions and COVID-19 mitigation strategies, such as vaccination, remains high in many African countries. A recent study examining vaccine sentiments in African societies found that out of more than 30,000 participants, only 48% of them reported acceptance of the COVID-19 vaccine [50]. The significant number of families unwilling to vaccinate themselves and their children could lead to increased infectivity rates, severe diseases, hospitalization, and deaths in many communities. As a result, vaccine hesitancy has become a more frequently discussed topic of conversation among researchers in Africa, who are raising awareness about vaccine safety and effectiveness [51]. The use of social media analytics, engagement-driven platforms—for example, social media and social listening tools—enable the collection of heterogenous data that inform the creation and amplification of tailored messaging, which target populations that are vaccine-hesitant and with limited access to care. This has the potential to tackle the infodemic crises, mis- or disinformation, vaccine hesitancy, and vaccine inequity.

Comprehensive educational programs focused on preventive health care and COVID-19 are becoming more widely accepted across Africa, creating opportunities for the scaling up of digital health technologies to facilitate digital literacy and precision health education or promotion. Training, capacity building, continuing professional development, and awareness or educational campaigns to bolster digital proficiency for both health care professionals and decision-makers can be achieved through structured web-based learning (e-learning) and mHealth technologies that provide access to comprehensive training programs in resource-limited areas that do not have access to the required personnel. With proper training, health care professionals are adequately equipped to assess patients and accurately report collected data via digital health systems. Importantly, communities and end users should be fully engaged and educated on digital health literacy, health information–seeking behaviors, vaccine safety or efficacy, etc. Institutions of learning (e.g., schools, colleges, and universities) should be adequately funded and equipped to offer courses or programs on big data analytics, digital health, and artificial intelligence [46]. Further, participation and infrastructure from private sector stakeholders (e.g., information and communication technology companies) should be enlisted to facilitate efficient operationalization and optimization of digital health initiatives.

Notably, the unique socio-multicultural make-up of African societies has led to misrepresentation and misunderstanding of inequality measures. The conventional and monocultural viewpoints of evaluating Social Determinants of Health (SDoH) indicators fall short of fully encompassing inequality measures within African societies. Therefore, the application of inequality measures from western countries to less industrialized countries may not generally represent the same outcomes [47,52]. Implementation of digital health could undoubtedly improve monitoring and reporting of SDoH indicators. This could accelerate attaining the sustainable development goals and universal health coverage and facilitate a more efficient response to future outbreaks and pandemics. It is pertinent that the most vulnerable within societies and hardest-to-reach regions are taken into account during policy- or decision-making and implementation of digital health systems architecture.

The African continent should have a formidable infrastructure moving into the mid-2020s with increasing opportunities to implement digital health solutions. Overall, these efforts toward digital health scalability ought to be government-driven, government-funded, nationally owned, sustainable, coordinated, ethical, cost-effective, and socioculturally aware. Accordingly, it is conceivable that Africans will be much better prepared to tackle and control future disease outbreaks, epidemics, and pandemics.

## Acknowledgments

This study is partially supported by Grant# 1R37CA234119-01A1 from National Cancer Institute (NCI).

## Authors' Contributions

OAO conceptualized and supervised the study and drafted, reviewed, and edited the manuscript. BW conceptualized the study and drafted the manuscript. CAM drafted the manuscript. AS-N drafted, reviewed, and edited the manuscript, supervised the study, and acquired the funding.

## Conflicts of Interest

None declared.

## Abbreviations

**Africa CDC:** Africa Centres for Disease Control and Prevention
**DHT:** digital health technology
**EHR:** electronic health record
**FETP:** Field Epidemiology Training Program
**mHealth:** mobile health
**SDoH:** Social determinants of Health

**WHO:** World Health Organization